\documentstyle[prd,aps,preprint,floats,epsfig]{revtex}

\tighten

\begin{document}
\draft
 
\pagestyle{empty}

\preprint{
\noindent
\today \\
\hfill
\begin{minipage}[t]{3in}
\begin{flushright}
LBL--51250 \\
\end{flushright}
\end{minipage}
}

\title{$\Lambda$ hyperon as helicity analyzer of $s$ quark 
in $B$ decay}

\author{Mahiko Suzuki}

\address{
Department of Physics and Lawrence Berkeley National Laboratory\\
University of California, Berkeley, California 94720
}
 
\maketitle

\begin{abstract}

 We explore how well one can probe the $s$ quark chirality of the 
fundamental weak interaction of nonleptonic $B$ decay using the 
spin-analyzing property of the $\Lambda$ hyperon. We present the 
prediction of the Standard Model as quantitatively as possible 
in a perturbative QCD picture avoiding detailed form-factor 
calculation involving quark mass corrections. A clean test of 
chirality will be possible with $\overline{B}\to\Lambda X$.

\end{abstract}
\pagestyle{plain}
\narrowtext

\setcounter{footnote}{0}

\section{Introduction}

    In the Standard Model only the left-chiral quarks enter
the fundamental weak interaction. Short-distance loop
corrections generate the penguin-type transition $b\to
s_L(d_L)$ as an effective decay interaction. This chiral
property leads to simple testable constraints on light meson
helicities in final states of $B$ decay if the final-state
interaction is perturbative\cite{Suzuki}. In order to determine 
chirality of weak interaction, we must measure sign of a meson 
helicity, $h=+1$ or $h=-1$, more than just transverse 
($h=\pm 1$) or longitudinal ($h=0$). Spin analysis 
of $B\to 1^-1^-$ through angular correlations was formulated 
by Dighe {\em et al}\cite{Dighe} and the data were analyzed
for $B\to J/\psi K^*\to l^+l^-K\pi$\cite{J/psi1,J/psi2}.
Without lepton spin measurement, however, this analysis is not
capable of distinguishing between $h=+1$ and $h=-1$ since it
leaves a twofold ambiguity in the transverse helicities\cite{S2}
A complete helicity determination can be achieved only with
spin and angular correlations, as shown in a more general 
formulation by Chiang and Wolfenstein\cite{Chiang}.

It has been argued that the penguin transition such as
$b\to g^*s$ and $\gamma^{(*)}s$ is more sensitive to a
nonstandard weak interaction than the tree interaction. 
The interaction $\overline{b}\to \gamma\overline{s}$ leads 
to $B\to\gamma K^*$ among others. In the Standard Model, 
$K^*(=\overline{s_L}q)$ ought to be produced in the helicity 
$+1$ state in this decay in the limit of $m_s=0$ and zero 
transverse momentum. If we wish to prove experimentally that 
$\gamma$ and $K^*$ are emitted with helicity $+1$ as predicted, 
we have to make a demanding measurement of lepton spin in 
$B\to\gamma^*K^*\to l^+l^-K\pi$\cite{gamma}. 
An alternative proposal was made to study 
the angular distribution of the process $B\to\gamma K_1\to
\gamma K\pi\pi$\cite{Gronau}. In this case the strong phase 
difference due to the overlapping resonances $\rho K$ and 
$K^*\pi$ of $K\pi\pi$ will allow us to obtain the $K_1$ spin 
information. Experimental efforts are being made on 
$B\to\gamma K\pi\pi$\cite{exp}.

Determination of the helicity sign is difficult in the cascade 
decays so far considered since parity is conserved in the 
second step of decay. If an intermediate particle of nonzero 
spin decays into final particles with a parity violating 
interaction, it is easy to determine the helicity sign through 
the spin-angular correlation $\langle{\bf s}\cdot{\bf p}\rangle$. 
The $B$ decay into a $\Lambda$ hyperon will provide us with such a
an opportunity since $\Lambda$ decays into $\pi N$ with the 
well-measured large parity asymmetry.  Furthermore, according to 
hadron spectroscopy, $\Lambda$ has the unique property that its 
spin is equal to the spin of the valence $s$ quark. Consequently 
the $s$ quark helicity can be determined by measuring the 
$\Lambda$ spin through a simple angular correlation of 
$\Lambda\to\pi N$. To probe a nonstandard weak interaction, 
therefore, it makes sense to explore the $s$-quark chirality 
in the QCD penguin interaction with $\Lambda$ as a spin analyzer.
 
\section{$\Lambda$ helicity versus strange quark helicity}

Ground-state baryons are made of three valence quarks totally 
in $s$-wave. Inside $\Lambda$ the $u$ and $d$ quarks form a 
spin singlet. As it is well known, therefore, the $\Lambda$ 
spin is made entirely of the $s$-quark spin in the static quark 
model. Boosting it to a moving frame, the helicity of 
$\Lambda$ is equal to that of the $s$ quark. The boost does not 
generate a new helicity component $l_z$ from the orbital motion 
since the distribution of $s$ quark is spherically symmetric 
inside $\Lambda$. When this $s$ quark comes directly from weak 
interaction, the handedness of $\Lambda$ tells us of the 
$s$-quark chirality in weak interaction. The $s$ quark can 
also be generated through pair production by gluons.  

The first task is to determine the helicity content of $s_L$
in flight that determines the $\Lambda$ helicity. If we could 
ignore the $s$ quark mass, the whole story would be trivial. 
In the real world, however, the $s$ quark carries mass and 
transverse momentum. When an $s$ quark is produced with
momentum ${\bf p}_s$ from the left-chiral field 
$\overline{s_L}$, projection of a plane wave shows that 
it is in the helicity $h=\pm\frac{1}{2}$ states with the 
amplitude ratio of
\begin{equation}
       \frac{A_{+\frac{1}{2}}}{A_{-\frac{1}{2}}}
 = \frac{E_s+m_s-|{\bf p}_s|}{E_s+m_s+|{\bf p}_s|}
 = \frac{m_s}{E_s + |{\bf p}_s|}, \label{ratio}
\end{equation}
where $E_s = \sqrt{m_s^2+{\bf p}_s^2}$. Inside $\Lambda$, the 
transverse quark momentum is part of the constituent quark 
mass. Therefore it is appropriate to replace $m_s$ with the 
constituent mass $M_s$ when we later express $A_{+1/2}/A_{-1/2}$ 
of $\Lambda$ in $|{\bf p}_{\Lambda}|$. Short-distance QCD 
interactions can alter the ratio of 
Eq.(\ref{ratio}) by $O(\alpha_s M_s/\pi E_s)$ for $M_s\ll E_s$. 
This is relatively a small correction even for only moderately 
fast $\Lambda$; for instance, $\alpha_s M_s/\pi E_s \simeq 0.08$ 
for $\alpha_s = \frac{1}{2}$ and $\gamma (= E_s/M_s) = 2$ in the
$\overline{B}$ rest frame. Our argument would obviously break 
down when a long-distance interaction plays a role in $\Lambda$ 
production, for instance, when $\Lambda$ is produced by 
$\overline{B}\to\Sigma(1385)X\to\pi\Lambda X$. It is easy to 
remove such a $\Lambda$ resonance band, if any.

 Which reference frame should we choose for Eq.(\ref{ratio})?
In a fast-moving frame of $\overline{B}$ where the $s$ quark moves 
even faster, the $\overline{s_L}$ field would produce the $s$ 
quark almost entirely in $h=-\frac{1}{2}$. Boosting it back to 
the $\overline{B}$ rest frame, one might reason that the $s$ quark 
and therefore the $\Lambda$ hyperon are almost 100\% in the 
$h=-\frac{1}{2}$ state since helicity is invariant under the
Lorentz boost. On the other hand, if one made the helicity 
projection in the $s$-quark rest frame, the $\overline{s_L}$ 
field would lead to $h=\pm\frac{1}{2}$ in a 50-50 probability. 
This apparent frame dependence is not physical, of course. The 
reason is that Eq.(\ref{ratio}) is only a projection of the 
$s$-quark plane-wave by $1-\gamma_5$. The complete decay amplitude 
is frame independent after the remainder of matrix element is 
combined. To see the point, we show the frame independence
for the hadronic two-body decay $\overline{B}\to 
\Lambda({\bf p}h)\overline{p}({\bf p}'h')$ instead of a quark 
process. The decay amplitude is of the form
$\overline{u}_{{\bf p}h}(A+B\gamma_5)v_{{\bf p}'h'}$.
In the two-component helicity spinors, it can be expressed as 
\begin{equation}
  \chi^{\dagger}_{h}\biggl(-A\sigma_3\sinh\frac{\eta-\eta'}{2}
   + B\cosh\frac{\eta-\eta'}{2}\biggr)\chi_{h'},
\end{equation}
where $\eta(\geq 0)$ and $\eta'(\leq 0)$ are the rapidities ($
\tanh\eta=p/E$) of $\Lambda$ and $\overline{p}$, respectively. 
(For $\eta >\eta'\geq 0$, $\chi_{h'}\to\chi_{-h'}$). Therefore 
the ratio of two $\Lambda$-helicity amplitudes is given by
\begin{equation}
     \frac{A_{+\frac{1}{2}}}{A_{-\frac{1}{2}}}=
       \frac{B-A\tanh\frac{1}{2}(\eta-\eta')}{
             B+A\tanh\frac{1}{2}(\eta-\eta')}\;\;\;
      {\rm for}\;\; B\to\Lambda\overline{p}. \label{ratioA}
\end{equation}
This is manifestly frame independent since the rapidity 
difference is invariant under the longitudinal Lorentz boost. 
In the $\Lambda$ rest frame, for instance, the $\Lambda$ 
helicity is determined by the helicity of the fast 
moving $\overline{p}$ through overall angular momentum 
conservation. In fact, Eq.(\ref{ratioA}) holds more generally. 
For $\overline{B}\to \Lambda X$, we can lump $X$ together into 
a single spinor of general spin and write the decay amplitude 
as $\overline{u}_{{\bf p}h}(A+B\gamma_5)p_{\mu}p_{\nu}\cdots 
v^{\mu\nu\cdots}_{{\bf p}'h'}$ since $\gamma_{\mu}$, 
$\gamma_{\mu}\gamma_5$, and $\sigma_{\mu\nu}$ either reduce to
$1$ and $\gamma_5$ or drop out by the Dirac equation or by the
subsidiary conditions on $v^{\mu\nu\cdots}_{{\bf p}'h'}$. Then 
Eq.(\ref{ratioA}) is reproduced. The boost invariance of  
Eq.(\ref{ratioA}) is nothing more than Lorentz invariance of the 
entire decay amplitude. The frame-independence argument holds 
likewise at the quark level though the individual emission and 
absorption vertices of quarks and gluons are not scalars nor 
pseudoscalars. We shall use the form of Eq.(\ref{ratioA}) as 
a guide to make our choice of frame.

The choice of frame would not be an issue if the $s$-quark mass
were zero ($A_{+1/2}/A_{-1/2}\to 0$). If we approach the 
problem by computing weak decay form factors, we have to know 
all relevant form factors including the quark-mass and 
transverse-momentum corrections. The numerator of 
Eq.(\ref{ratio}) is such a correction term arising 
from difference of two large form factors. 
In perturbative QCD and the light-cone description of hadrons, 
we can obtain it, in principle, by computing higher twist terms
with spin-dependent quark distribution functions. In practice, 
however, it is difficult to reach quantitatively reliable 
answers even for the $B$ decay into two mesons.\footnote{
Spin dependence was studied extensively by theorists
for $B\to J/\psi K^*$. In this decay the dominant contribution 
to the $h=\pm 1$ final helicities arises from $m_c\neq 0$, not
from $M_s\neq 0$. While theoretical predictions have converged
to the experimental values\cite{J/psi1,J/psi2} with time, one  
sees how widely theoretical predictions used to spread when
no data were available.\cite{J/psi1}}
No factorization limit exists for $\overline{B}\to\Lambda X$.
Giving up computing the $O(M_q)$ terms of form factors, we shall 
present alternative semiquantitative results by stretching the 
perturbative QCD picture to the limit.

There is one basic problem about quark rapidities. While we 
can determine a hadron rapidity directly from experiment, a quark 
rapidity inside a hadron has a continuous distribution which we 
do not know precisely. We circumvent 
this problem by introducing an approximation. Since $\Lambda$ is
an $s$-wave ground state in the rest frame, one reasonable 
approximation is to substitute $E_s$ and ${\bf p}_s$ with their 
average values inside $\Lambda$:
\begin{eqnarray}
    E_s &\to& \langle E_s\rangle
           \simeq\frac{M_s}{M_u+M_d+M_s}E_{\Lambda}, \nonumber \\ 
   {\bf p}_s&\to&\langle{\bf p}_s\rangle
           \simeq\frac{M_s}{M_u+M_d+M_s}{\bf p}_{\Lambda},
                         \label{ave}
\end{eqnarray}
where $M_s/(M_u+M_d+M_s)\simeq 0.45$. Eq.(\ref{ave}) means
that the $s$ quark moves on average with the same Lorentz factor 
$\gamma$ as $\Lambda$ does in a moving frame.

Now we choose the frame in which the helicity amplitude ratio 
is determined with Eqs.(\ref{ratio}) and (\ref{ave}). The rest 
frame of $\overline{B}$ may come to our mind as an obvious 
choice. But a better alternative is the rest frame of $X$ for 
$\overline{B}\to\Lambda X$. In this frame, rapidity $\eta'=0$ in 
Eq.(\ref{ratioA}) or its generalization to $\overline{B}\to\Lambda 
X$ so that the helicity ratio depends only on $\eta$. If we want 
to express the ratio in terms of the energy-momentum of $\Lambda$ 
quark alone without involving $X$, therefore, we should choose 
the rest frame of $X$. The helicity amplitude ratio is given by
Eq.(\ref{ratio}) with the energy-momentum $E'_{\Lambda}$ and 
${\bf p}'_{\Lambda}$ of the $X$ rest frame:
\begin{equation}
  \frac{A_{+\frac{1}{2}}}{A_{-\frac{1}{2}}}
    = \frac{E'_{\Lambda}+ m_{\Lambda}-|{\bf p}'_{\Lambda}|}{
    E'_{\Lambda}+ m_{\Lambda}+|{\bf p}'_{\Lambda}|}
    =\frac{m_{\Lambda}}{E'_{\Lambda}+|{\bf p}'_{\Lambda}|},
     \label{ratio1}
\end{equation}
where $M_u+M_d+M_s\simeq m_{\Lambda}$ has been used for the 
constituent quark masses. We would obtain Eq.(\ref{ratio1}) if 
we simply project the $\Lambda$ field onto 
$(1-\gamma_5)\psi_{\Lambda}$.\footnote{
That is what we expect since the $\Lambda$ spin is equal 
to the $s$-quark spin, and $\Lambda$ and $s$ move with the 
same Lorentz factor. Such a projection is obviously not valid 
for the decay $\Lambda\to \pi N$ since the process involves 
very strong nonperturbative effects, the long-distance 
$\Delta I = \frac{1}{2}$ enhancement.}
The ratio of Eq.(\ref{ratio1}) can be expressed in terms of
the quantities in $B$ rest frame as
\begin{equation}
      \frac{A_{+\frac{1}{2}}}{A_{-\frac{1}{2}}} =
    \biggl(\frac{m_{\Lambda}}{E_{\Lambda}+|{\bf p}|}\biggr)
    \biggl(\frac{m_X}{E_X+|{\bf p}|}\biggr)
      \equiv \delta(E_{\Lambda}), 
                             \label{ratio2}
\end{equation}
where $|{\bf p}|^2 = E_{\Lambda}^2-m_{\Lambda}^2$, $E_X=m_B 
- E_{\Lambda}$, and $m_X^2=m_B^2-2m_BE_{\Lambda}+m_{\Lambda}^2$.

\section{Tree and penguin interactions}

Even in the Standard Model, a hard $s_R$ can contribute to 
formation of $\Lambda$ through the penguin interaction. To test 
the Standard Model with the $\Lambda$ helicity, therefore, we 
need to know the $s_R$ contribution of the penguin interaction. 
For the purpose of separating this $s_R$ from $s_L$, we 
parametrize relative importance of the penguin interaction 
to the tree interaction by
\begin{equation}
 p=\frac{d\Gamma_{\rm penguin}/dE_{\Lambda}}{
         d\Gamma_{\rm tree}/dE_{\Lambda}},
\end{equation}
where $p$ is generally a function of $E_{\Lambda}$.
Even in the two-body meson decays, $B\to K\pi$ and
$B\to\pi\pi$, the relative weight of the two types of 
interactions has not been well determined from experiment. 
It is generally agreed among theorists that when $X$ has net 
strangeness zero, the dominant interaction is the penguin 
interaction, {\em i.e.}, $p>1$ though the tree interaction 
may not be totally negligible. Theoretical uncertainties are 
smaller for inclusive decays, but the limited accuracy of 
$V_{ub}$ at present still makes it difficult to determine 
the value of $p$ with certainty; $p\approx 3-10$ for 
$|V_{ub}|= 0.0025-0.0048$. Fortunately, however, the Standard 
Model prediction turns out to be insensitive to the value of $p$. 
When $X$ has one unit of net strangeness ($X_{\overline{s}}$), 
the penguin interaction $(\overline{b}d)(\overline{s}s)$ and the 
tree interaction $(\overline{b}u)(\overline{u}d)$ followed by 
$\overline{u}u\to\overline{s}s$ are responsible for the decay. 
As for the relative strength between $X$ of $X_{\overline{s}}$, 
the smallness of $|V_{td}/V_{ts}|$ and $|V_{ub}|$ suppresses 
$X_{\overline{s}}$ relative to nonstrange $X$. In the two-body 
meson decays, this statement suggests $B(B\to K\overline{K},
\pi\pi)\ll B(B\to K\pi)$; experimentally\cite{kpi,PDG}, 
$B(B\to K^+\pi^-)/B(B\to\pi^+\pi^-)\simeq 3.3\pm 0.5$ and 
$B(B\to K^+\pi^-)/B(B\to K^+K^-) >15$. ItTherefore, its 
reasonable to expect that net strangeness of $X$ is most often 
zero in $\overline{B}\to\Lambda X$. We proceed with the 
approximation that $X$ has net strangeness zero. When a value 
of $p$ is relevant, we choose $p\gg 1$ to reflect the penguin 
dominance in $\overline{B}\to\Lambda X$; more specifically,
in the range of
\begin{equation}
             p \approx 6\pm 3. 
\end{equation} 

\section{Helicity ratio from angular asymmetry}

The angular distribution of the cascade decay $\overline{B}\to
\Lambda X\to \pi^- pX$ can be written in the form
\begin{equation}
  \frac{d^2\Gamma}{dE_{\Lambda}d\cos\theta}
     =\frac{1}{2}\frac{d\Gamma}{dE_{\Lambda}}
       (1+\overline{\alpha}\cos\theta),
                            \label{asym}
\end{equation}
where $\theta$ is defined as the emission angle of proton in the 
$\Lambda$ rest frame that is measured from the direction of the 
$\Lambda$ momentum ${\bf p}$ of the $\overline{B}$ rest frame. 
Knowing the ratio $\delta(E_{\Lambda})$ for $\Lambda$ helicity 
from Eq.(\ref{ratio2}), we can relate $\overline{\alpha}$ to the 
$\Lambda$ decay asymmetry parameter $\alpha_{\Lambda}
= 0.642\pm 0.013$\cite{PDG}. Taking account of coexistence of the
tree and penguin interactions, we can express $\overline{\alpha}$ 
in terms of $\alpha_{\Lambda}$ by counting left and 
right-chiral $s$ fields in $(\overline{b}u_L)(\overline{u_L}s_L)$ 
of the tree interaction and $(\overline{b_L}s_L)(\overline{q_L}q_L
+\overline{q_R}q_R)$ ($q=u,d,s$) of the QCD penguin interaction:
\begin{equation}
 \overline{\alpha} = 
       - \biggl(\frac{1+\frac{4}{5}p}{1+p}\biggr)
   \biggl(\frac{1-\delta(E_{\Lambda})^2}{
                1+\delta(E_{\Lambda})^2}\biggr) \alpha_{\Lambda}.
                         \label{alpha}
\end{equation}
where we have ignored the electroweak penguin interaction, 
the interference between the tree and the QCD penguin (an
approximation better for inclusive than exclusive decays), and
the phase space difference between $u/d$ and $s$. The first 
factor varies only from $0.9$ to $0.8$ over 
the range of $p$ from 1 to $\infty$. The second factor in the 
right-hand side of Eq.(\ref{alpha}) is practically unity over 
a wide range of $E_{\Lambda}$ except near the low energy end. 
The asymmetry $\overline{\alpha}$ approaches zero in the slow 
limit of $\Lambda$ ($\delta(m_{\Lambda})=1$). This limiting 
value is a kinematical constraint since no preferential direction 
exists in space in this limit where all momenta are either zero or 
integrated over. The asymmetry $\overline{\alpha}$ moves rapidly 
from 0 to about $-0.4$ at $E_{\Lambda}= 1.5$ GeV and then 
approaches slowly $-\alpha_{\Lambda}$ up to the factor 
$(1+\frac{4}{5}p)/(1+p)\simeq 1$. The negative $\overline{\alpha}$
means the $h=-\frac{1}{2}$ dominance for the $s$ quark from 
the $\overline{s_L}$ field.

If a nonstandard interaction generates the QCD penguin interaction
$\overline{b}(1-\kappa\gamma_5)s(\overline{q}q)$, the asymmetry is
\begin{equation}
 \overline{\alpha}= 
    -\biggl(\frac{1+8\kappa p/5(1+\kappa^2)}{1+p}\biggr)
    \biggl(\frac{1-\delta(E_{\Lambda})^2}{
                 1+\delta(E_{\Lambda})^2}\biggr)\alpha_{\Lambda}.
\end{equation}
If a significant amount of $s_R$ mixes in the QCD penguin
interaction, $\overline{\alpha}$ would be close to zero or even 
positive in contrast to the negative values predicted for 
the Standard Model. While a precise value of asymmetry  
depends on the value of $p$, $\overline{\alpha}$ would show a 
marked departure from the prediction of the Standard Model in 
this case. This is the helicity test that we propose in 
this paper. 

Plotted in Figure 1 is the asymmetry $\overline{\alpha}$ expected 
for $\overline{B}\to\Lambda X$ in the Standard Model. The curve
is plotted for $p=3$ so that it is subject to a small uncertainty 
of $\pm 2\%$ (for $p$ = 3 to 9). The perturbative QCD correction 
below $m_b$ of $O(\alpha_sM_s/\pi E_s)$ is the main uncertainty, 
which is $O(10\%)$ at the higher half of the $E_{\Lambda}$. 
While the QCD correction is process dependent, a deviation of 
20\% or more from the curve in Figure 1 will be a clear warning 
sign of a wrong helicity $s$ quark in weak interaction, or else, 
breakdown of perturbative QCD in final-state interactions.

\noindent
\begin{figure}[h]
\epsfig{file=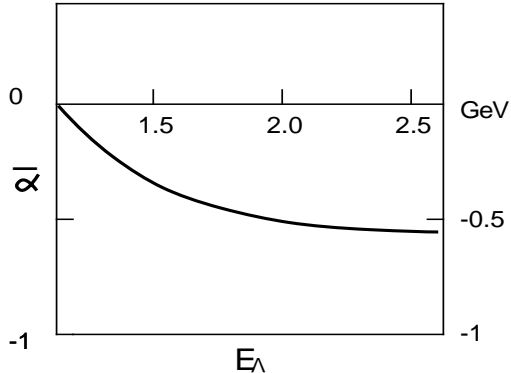,width=7cm,height=5cm}
\caption{The asymmetry $\overline{\alpha}$ in the Standard Model
plotted against the $\Lambda$ energy in the $B$ rest frame
(for $p=3$).
\label{fig:1}}
\end{figure}

A comment is in order on the background from the
$b\to c$ transition. The interaction $b\to c_L\overline{u_L}
(\overline{c_L})s_L$ can only lower $\overline{\alpha}$, that 
is, increase the magnitude of $|\overline{\alpha}|$ since the 
$s$ quark field is left-chiral. The final state of the lowest
mass for $b\to c_L\overline{u_L}s_L$ is $\Lambda D\overline{N}$ 
($5.26$ GeV vs $m_B$= 5.28 GeV). The phase space suppression 
virtually eliminates this mode. The cascade weak decay 
$\overline{B}\to\Lambda_c X\to\Lambda X'$ through 
$b\to c_L\overline{u_L}d_L
\to s_L\overline{u_L}d_L\overline{u_L}d_L$ is more favorable 
in phase space and in the quark mixing. The 
branching fraction of $\Lambda_c\to\pi\Lambda$ is about $1\%$ 
and the inclusive branching to $\Lambda X$ is $\sim 10\%$. If 
the decay process $c_L\to s_L\overline{d_L}{u_L}$ occurs 
perturbatively, the final $s$ quark is left-chiral so that it 
tends to lower $\overline{\alpha}$. In any way we can separate 
the $\Lambda_c$ band, if necessary. Therefore the $b\to c$ 
transition will not pose a problem.

\section{Remark and conclusion}

Analysis in the $B$ decay modes feeding $\Lambda$ is still at
an early stage. Only an upper bound has been set on the
branching to two-body baryonic channels, {\em e.g.},
$B(B^+\to p\overline{\Lambda})<2.2\times 10^{-6}$\cite{Belle0}.
However, the decay into three bodies, $B^{\pm}\to p\overline{p}
K^{\pm}$\cite{Belle2} has been observed with the branching 
fraction of $(4.3^{+1.1}_{-0.9}\pm 0.5)\times 10^{-6}$. 
The decay $B\to p\overline{\Lambda}\pi$ and the conjugate occur
presumably at the same level of branching fraction. The
inclusive decay events $\overline{B}\to\Lambda X$ will be seen
abundantly in near future.

 To conclude, measurement of the $\Lambda$ decay asymmetry in 
$\overline{B}\to\Lambda X$ is a sensible test to probe the 
chirality structure of the fundamental weak interaction. 
It will test whether the QCD penguin interaction possibly 
contains a nonstandard term such as $(\overline{b}s_R)
(\overline{q}q)$ or not. Our numerical predictions contain 
inevitable uncertainties as we have delineated. Nonetheless, 
experimental determination of helicity with $\Lambda\to\pi^- p$ 
will be cleaner than that with the radiative $B$ decay. We believe 
that the decay $\overline{B}\to\Lambda X$ will be competitive 
with, if not superior to, $B\to\gamma^{(*)}X$ in testing the 
chiral structure of the penguin interaction at B factories. 
It also has an advantage over $\Lambda_b\to\gamma\Lambda$ at 
hadron colliders where the $\Lambda_b$ polarization introduces 
another uncertainty.

\acknowledgements

This work was supported in part by the Director, Office of
Science, Office of High Energy and Nuclear Physics, Division
of High Energy Physics, of the U.S. Department of Energy under
contract DE-AC03-76SF00098 and in part by the National Science
Foundation under grant PHY-0098840.


 
\end{document}